\documentclass[a4paper,12pt]{article}

\usepackage{authblk}

\usepackage{fullpage}
\usepackage{graphicx}
\usepackage{subfigure}
\usepackage{epsfig}
\usepackage{dcolumn}
\usepackage{bm}

\usepackage{amsmath,amssymb,cite}
\begin{document}

\title{A multipurpose information engine that can go beyond the
Carnot limit}
\author{Shubhashis Rana\footnote{email: shubhashis.rana@gmail.com},  ~and A. M. Jayannavar\footnote{email: jayan@iopb.res.in}}
\affil{Institute of Physics, Sachivalaya Marg, Bhubaneswar - 751005, India}
\date{}

\maketitle{}





 \begin{abstract}
  
 Motivated by the recent work  by  Mandal and  Jarzynski  on autonomous Maxwell  demon information 
 engine, we have extended their model by introducing two different heat baths. The system (demon) is coupled 
 to a memory register (tape) and a work source. The performance of the system 
 depends on the interplay between these two sources along with the heat baths.
 We have found the system can act as an engine, refrigerator or an eraser. Even the 
 combination of any two is possible in some parameter space. We have achieved the
 efficiency of the engine is greater than Carnot limit. The coefficient of performance
 of refrigerator also achieves larger than Carnot limit.
 \end{abstract}
 
  \hspace{0.2cm}\small PACS: 05.40.Ca, 05.70.Ln, 89.70.Cf

 \normalsize
  \hspace{0.2cm}Keywords: Non-equilibrium statistical Mechanics, Thermodynamics of information processing, Maxwell demon


\newcommand{\nwc}{\newcommand}
\nwc{\vs}{\vspace}
\nwc{\hs}{\hspace}
\nwc{\la}{\langle}
\nwc{\ra}{\rangle}
\nwc{\lw}{\linewidth}
\nwc{\nn}{\nonumber}

\nwc{\pd}[2]{\frac{\partial #1}{\partial #2}}
\nwc{\zprl}[3]{#3~Phys. Rev. Lett. ~{\bf #1},~#2}
\nwc{\zpre}[3]{#3~Phys. Rev. E ~{\bf #1},~#2}
\nwc{\zpra}[3]{#3~Phys. Rev. A ~{\bf #1},~#2}
\nwc{\zjsm}[3]{#3~J. Stat. Mech. ~{\bf #1},~#2}
\nwc{\zepjb}[3]{#3~Eur. Phys. J. B ~{\bf #1},~#2}
\nwc{\zrmp}[3]{#3~Rev. Mod. Phys. ~{\bf #1},~#2}
\nwc{\zepl}[3]{#3~Europhys. Lett. ~{\bf #1},~#2}
\nwc{\zjsp}[3]{#3~J. Stat. Phys. ~{\bf #1},~#2}
\nwc{\zptps}[3]{#3~Prog. Theor. Phys. Suppl. ~{\bf #1},~#2}
\nwc{\zpt}[3]{#3~Physics Today ~{\bf #1},~#2}
\nwc{\zap}[3]{#3~Adv. Phys. ~{\bf #1},~#2}
\nwc{\zjpcm}[3]{#3~J. Phys. Condens. Matter ~{\bf #1},~#2}
\nwc{\zjpa}[3]{#3~J. Phys. A: Math theor  ~{\bf #1},~#2}

 \section{Introduction}
  
 Since the advent of Maxwell demon 150 years ago, the study of relationship between 
 thermodynamics and information has attracted much attention because it concerns 
 the foundation of second law of thermodynamics \cite{max71,szi29,lan61,pen70,
 ben82,ben85,zur89,leff03,ved09,man12,man13,bar13,bar14,def13,str13,hop14,hor13,esp12,peng16,jar13}.
 Maxwell demon is formulated as an information processing device that performs measurement including memory
 register and feedback at the level of thermal fluctuations. This demon refers to 
 any general situation in which rectification of microscopic fluctuations decrease 
 thermodynamic entropy. To preserve the second law from violating, information to 
 memory register plays an essential role. Information entropy of the memory 
 register compensates the decrease of the thermodynamic entropy of the device\cite{jar13,
 sag09,par15}.  Recent studies have revealed that information content and thermodynamic variables to be
 treated on equal footings (information thermodynamics)\cite{ved09,jar13,par15}. This relationship  has attracted 
 much interest recently due to advances of nano-technology which have enabled to access
 atomic-scale objects in a controllable manner. Experimentally Maxwell demon and Szilard 
 engine have been realized in laboratory\cite{toy10,kos14,ber11}. This has allowed verification of several 
 fundamental issues regarding thermodynamics of information and physical 
 nature of information.

 Recently Mandal and Jarzynski (MJ) have introduced an autonomous maxwell demon model \cite{man12} for heat engine. Apart from heat
 reservoir the model consists of three parts such as work source, the device and information reservoir. A mass 
 that can be raised or lowered against gravity is used as a work source while the device  
 operates in cycle and affects the transfer of energy among other subsystems.  An information reservoir
 is a system that exchanges information but not energy with the device. A frictionless tape 
 acts as an  information reservoir that preserves the information.
 The reading and writing of information  does not require any energy while passing
 through the demon. Depending upon the parameter, they have showed that the system sometime functions as an 
 engine or erasure or none of them(called dud state (MJ)).
 
In this paper, we have extended their model by introducing two reservoirs coupled to the system. 
The main motivation of our study is to find the effect of thermal
 bias along with information reservoir and the work reservoir in the performance of the demon. 
The system reaches an unique  steady state for any given set of parameters
(which will be described in detail below). The  dynamics are autonomous.
The performance of our model basically  depends on the  interplay
between different  forces: the gravitational pull on the mass and the changing the information 
content on the stream of bits written on the tape. The system can act as an engine by extracting 
work  on an average; an erasure by removing the information from the memory register (tape),
a refrigerator by absorbing heat from cold bath. Our system also exhibits  combination of any
of the two in its modes of operation. For example it simultaneously acts as i) an engine and a refrigerator,
ii) an engine and an eraser, iii) a refrigerator and an eraser. Our results are consistent with generalized
second law of thermodynamics. We will discuss  model and findings in detail below.
 
 \section{The Model:}
 
\begin{figure}[!ht]
\vspace{0.5cm}
\begin{center}

\includegraphics[width=12cm]{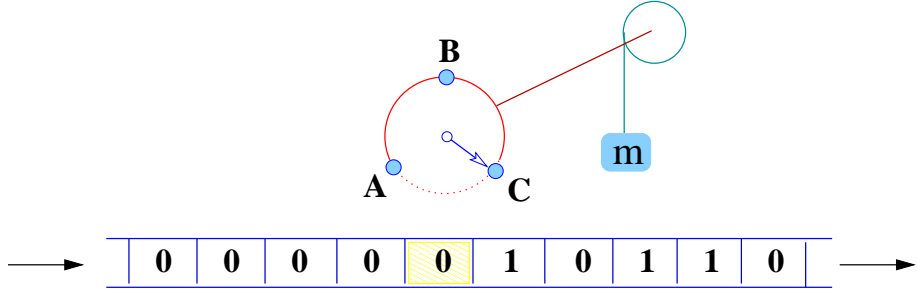}
\caption{ A schematic diagram of our model: The demon is a three state system which is coupled with a external load and 
two thermal reservoirs (not shown). A sequence of bits (tape) passes from left to right at a constant speed.
 The nearest bit interact with the demon. For positive load i.e,  $w>$ 0, the mass is lifted at an amount
 $\Delta h$ for every transition C $\rightarrow$ A while for every transition A $\rightarrow$ C the mass
 is lowered by same amount. However  $w <$ 0 the mass is connected to right side of the small circle, so the transition 
  C $\rightarrow$ A lowers the mass and  the transition A $\rightarrow$ C lift it up.}
\label{model1}
\end{center}
\end{figure}

\begin{figure}[!ht]
\vspace{0.5cm}
\begin{center}

\includegraphics[width=12cm]{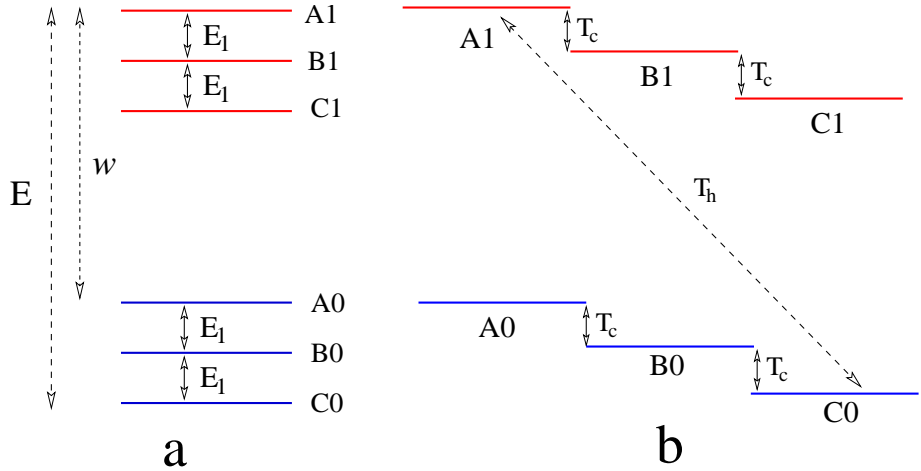}
\caption{ Possible six states depending on demon and the bit. (a) the difference between the energy levels 
between A and B is $E_1$ similarly the difference between B and C is also $E_1$. The energy absorbed/released
during any transition between C0 and A1 is denoted as E such that $E=w+2E_1$. (b) All the allowed transitions 
and corresponding bath where energy is exchanged.}
\label{model2}
\end{center}
\end{figure}

 We have extended the  autonomous Maxwell demon model by  MJ. To this end we have  considered
 similar  three state system (A, B, C) but these states have  different energy levels (unlike MJ), i.e., they are 
 not degenerate (see Fig.\ref{model1} and Fig.\ref{model2}). For simplicity we have 
 considered difference between the two successive energy levels to be same ($E_1$). Note that 
 minimum three states are needed to observe any directed rotation. The transition can 
 occur between A to B, B to C and vise versa spontaneously by exchanging heat from the cold bath 
 with temperature $T_c$ and internal energy of the system gets changed. However, the transition
 between A to C and vice versa is restricted and 
 depends on the value of the interacting bit written on the tape. The bit has two states 0 and 1.
 Hence the combined system  has 6 states. When the transition occur
 from C0 to A1, the bit state is changed from 0 to 1 and 
 vise versa. However, transition between C1 and A0 is not allowed.  Note that, during the transition between A and
 B, B and C the bit state is not changed. On the other hand, during the transition  from C0 to A1, 
  energy is absorbed from the hot bath at temperature $T_h$ by an amount $E$ while the system perform $w$
  amount of work by pulling a mass m to a height  $\Delta h$ by a frictionless pulley  in  a gravitational 
 force field $g$ ($w=mg\Delta h$). During this transition, the internal energy of the 
 demon is increased by $2E_1$. Using energy balance one can get 
 
 \begin{equation}
 E=w+2E_1.
\end{equation}

 \begin{figure}[!ht]
\vspace{0.5cm}
\begin{center}
 
\includegraphics[width=3cm]{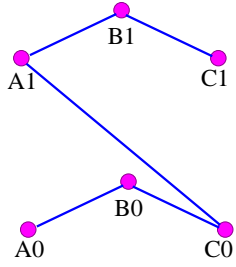}
\caption{ All the allowed transitions form a liner chain during  the evolution along with  the bit. }
\label{liner}
\end{center}
\end{figure}

The rate of transition between C0 and A1 is related by
 \begin{equation}
  \frac{R_{C0\rightarrow A1}}{R_{A1\rightarrow C0}}=e^{-E/T_h}.
 \end{equation}

 We set Boltzmann constant $k_B=1$. If we take $R_{C0\rightarrow A1}=1-\varepsilon$ and $R_{A1\rightarrow C0}=1+\varepsilon$,
 we get
 \begin{equation}
   \varepsilon=\tanh  \left(\frac{E}{2T_h}\right).
 \end{equation}
 This weight parameter $\varepsilon$ is bounded by $-1< \varepsilon <1$. The other 
 transitions occur due to connection to the cold bath $T_c$ and the rates are related by
 \begin{equation}
  \frac{R_{B0\rightarrow A0}}{R_{A0\rightarrow B0}} =\frac{R_{C0\rightarrow B0}}{R_{B0\rightarrow C0}}=e^{-E_1/T_c}
 =\frac{R_{B1\rightarrow A1}}{R_{A1\rightarrow B1}} =\frac{R_{C1\rightarrow B1}}{R_{B1\rightarrow C1}}.
  \end{equation}
 All the allowed transitions are shown in Fig.\ref{model2}b and they form a linear chain (Fig.\ref{liner}). The model 
 can not exhibit a directed rotation, only back and forth transitions are possible. To get any possible
 rotation we need to introduce the change of bit externally. We consider the tape is moving with
 a rate $\tau^{-1}$ from left to right. The system interacts  only with the 
 nearest bit (one bit at a time). Each bit is allowed to interact the   system with time $\tau$ before
 the next bit takes its place. The earlier bit with new information content move forward 
 along with the tape. If $\tau$ is very small the system hardly evolves 
 during this time. If $\tau$ is large, the system gets enough opportunity to evolve along with the bit.
 After  time $\tau$, the joint state may change and depends on the incoming bit. During this
 transition, actual state of the system (A, B or C) is not changed and hence it does not involve
 any energy cost. Take an example: if the demon state is B and the outgoing bit in state 1 then the 
 joint state will be B1. Now if the incoming bit is 0, then the joint state at the beginning of 
 the new cycle, will change to simply B0. Note that during this transition from B1 to B0 the internal 
 energy of the demon is not changed since demons actual state is fixed at B. 
 
 In our present study we  simulate this problem and analyze our findings.
 
 \section{Degenerate energy levels with a single thermal reservoir}
  First we set $E_1=0$ ($E=w$)  and $T_h=T_c$. This case correspond to the original  problem of MJ. Now 
 all the levels are degenerate.  Heat is exchanged only during the transition
 between A1 and C0 from a single bath with temperature T. Let $\delta $ denotes excess number of 0 in the incoming bit stream,
  
  \begin{equation}
   \delta =p(0)-p(1).
  \end{equation}
Here $p(0)$ and $p(1)$ represents probability of 0 and 1 respectively in the incoming bit stream and
they are normalized i.e., $p(0)+p(1)=1$. The Shannon entropy of the incoming bit stream 
(written on the memory register or tape) is given by 
\begin{equation}
 S=- \sum_i p_i\ln p_i=-p(0)\ln p(0)-p(1)\ln p(1).
\end{equation}
It represents the amount of disorder present per bit. We have ignored the correlations between 
the successive bits. $S$ quantifies the  information content in the incoming bit stream.
Consider the case when  all  incoming bits are 0; which implies $\delta=1$ and $S=0$.
After the interaction time $\tau$, if the joint state ends up at any of the following
states A1, B1 or C1, the outgoing bit is changed to 1. This implies during this time $\tau$, 
there is one excess number of upward transitions compared to the
downward transitions. If the final state is in A0, B0 or C0 the outgoing bit will be in 0 state.
For this situation, the number of upward transitions is balanced by the downward transitions. In long time
limit $t\gg \tau$, the system will attain the steady state along with the tape. The outgoing bit 
stream will be mixture of 0 and 1 with probability $p'(0)$ and $p'(1)$  respectively. Then the 
Shannon entropy of the  outgoing bit stream 
\begin{equation}
 S'=-p'(0)\ln p'(0)-p'(1)\ln p'(1).
\end{equation}
will be always positive. This means some information has been written on
the tape ($\Delta S=S'-S>0$). Every 1 in the outgoing bit stream represents $w$ work is extracted 
by pulling the mass at an amount $\Delta h$, taking the heat from the bath. Hence for $\delta=0$, on an average we have
extracted work by rectifying the noise from the single heat bath. This is possible because the entropy of the
memory register (tape) increases. Now if we consider entropy of the memory register and bath,
 it together follows the second law 
\begin{equation}
 \Delta S + \Delta S_B\ge0.
\end{equation}
$\Delta S_B$ represents entropy change of the bath. It may be noted that in steady state,
the  entropy change of the demon is zero  on average.

 For $-1<\delta <1$ the evolution of the combined system will depend on two forces: gravitational pull on the mass
 (depends on weight parameter $\varepsilon$) and the randomizations of bits ($\delta$) along with the duration 
 of the contact time of each bit $\tau$.  The average number of clockwise (CW) rotation is given by 
\begin{equation}
 \phi =p'(1)-p(1)=\frac{1}{2}(\delta-\delta')
\end{equation}
 where $\delta'=p'(0)-p'(1)$. Note that during each upward transition (CW rotation) E energy is absorbed from the
 heat bath and same amount of work extracted ($w=E$).  Then on average  
 \begin{equation}
  Q=-E\phi
  \label{def-Q}
 \end{equation}
 heat will dissipated into the bath.  As all the states of the demon is degenerate,
 one can easily find 
 \begin{equation}
  W=-w\phi=Q
 \end{equation}
  work will be done on the system on average. For $\phi>0$,   work is extracted 
  on average ($W<0$) and system acts as an engine.

 For $\delta=0$ the incoming bit stream contains maximum information and $S(0)=\ln 2$. Before the beginning of 
 each interval, the probability of finding the demon in upper state or lower state is equal
 ($p(0)=p(1)=\frac{1}{2}$). However  after the interaction, the system will  relax and
 lower state will have larger probability ($p'(0)>\frac{1}{2}>p'(1)$). This leads to the
 information content in the outgoing bits to be less than that of the incoming bits ($\Delta S<0$).
 Since $\phi<0$, we have $W>0$ (Eq.(\ref{def-Q})). This means work is done on the system to erase the information contained in the tape 
 and the system acts as an erasure.

 If all the incoming bits are 1 ($\delta=-1$), after the interaction we obtain mixture of 0 and 1 in the outgoing 
 bits. Every 0 in the outgoing bit stream represents a counter clockwise (CCW) rotation. Hence on average 
 $W>0$ and $\Delta S>0$. Here the system neither behaves as an engine nor an erasure and we call it dud.
 
 Till now we have considered $w>0$. The negative $w$ represents the load is connected to the
 right side of the pulley. It makes  for every transition C $\rightarrow$ A the mass is lowered while
 for every transition from A $\rightarrow$ C the mass is lifted up. As a result for $\delta = 1$ the 
 system acts as a dud while for $\delta=-1$ the system performs as an engine.
 
\begin{figure}[!ht]
 \vspace{0.5cm}
\begin{center}
 \includegraphics[width=15cm]{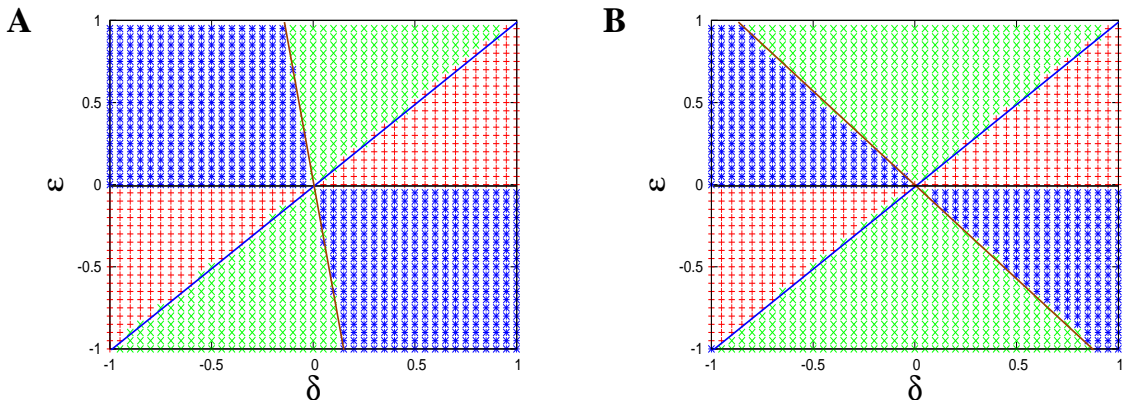}
\caption{A) Phase diagram when the demon is connected with single bath at temperature T=1.0 for 
$E_1$=0 and cycle time $\tau=1.0$. The phase diagram consist of three regions namely
i) Engine (Red plus), ii) Eraser (green cross), iii) Dud (blue star). Line 1 (Black line) 
separates engine and dud region, while line 2 (blue line) separates engine and erasure region and
finally line 3 (brown line) separates erasure and dud region.
B) It represents the phase diagram when $\tau=10$ and other parameters remains same.
}
\label{mandal1}
\end{center}
\end{figure}

 In Fig.\ref{mandal1}A we have plotted our numerically obtained  phase diagram for $\tau=1$. We find
 depending on the parameter $\delta$ and $\varepsilon$ the system can exhibit engine (red region), erasure (green region)
 or dud (blue region). This phase diagram exactly matches with the theoretical plots given by
  MJ. This also verifies the correctness of our simulation.

 In the phase diagram, we have found there are three boundary lines that separate the different operating regions.
 At $\varepsilon =0$, we have  set $E=0$ or $w=0$. That means during the CW or CCW rotation no work is done 
 on the system and  heat absorption is  zero and thus $W=0$ and $Q=0$. The second law becomes 
 $\Delta S>0$ since $\Delta S_B=\frac{Q}{T}=0$. We define this phase boundary where $W=0$  as 
 {\bf{line 1}} (shown as {\bf{black line}} in Fig.\ref{mandal1}) and it separates 
 the engine and the dud regions.
 
 At the phase boundary between  engine and the erasure region we have $W=0$ and $\Delta S=0$.
 We denote this boundary line as {\bf{line 2}} (as  denoted by {\bf{blue line}} in Fig.\ref{mandal1}).  
 It implies $ Q=0$ and total entropy production  is also zero. This can only happen when  probability distribution of bits does not 
 change during the interactions: $p'(0)=p(0)$ and $p'(1)=p(1)$. Which implies the initial 
 bit stream is such that the demon is already in equilibrium with the bath. $p(0)=\frac{e^{E/T}}{1+e^{E/T}}$
 and $p(1)=\frac{1}{1+e^{E/T}}$; hence line 2 occurs at  $\delta= p(0)-p(1)=\varepsilon$.  
 Note that here the demon operates  reversibly and total entropy production is zero. 
 
 The third boundary line which separates the erasure and the dud region is called
 as {\bf{line 3}} (as shown by {\bf{brown line}} in Fig.\ref{mandal1}). Here, $\Delta S=0$ but 
 we have observed from phase diagram $W>0$. The second law  reduces to $\Delta S_B>0$.
 This is only possible if the probability distribution of
 bits gets altered after the evolution along with the demon:  $p'(0)=p(1)$ and $p'(1)=p(0)$. 
 Line 2 and line 3 must cross each other at $\delta=0$ because for this case the two condition 
 together hold ($p(0)=p(1)=\frac{1}{2}$).  Putting $\delta=0$ one obtains $\varepsilon=0$ (as 
 this is the condition of line 2). Hence this crossing point must be on line 1 and three lines
 will meet at this point.

 For $\delta <0$ and   $\varepsilon \rightarrow 1$ the upper states are initially more populated ($p(1)>p(0)$). During the 
 evolution the system will  relax. The final state will depend on the relaxation time $\tau$. If $\tau$
is large it is possible to reach even $p'(0)>p(1)$. Then the system will act as an erasure. In Fig.\ref{mandal1}B
we have plotted the same phase diagram for $\tau=10$. It is  observed that as $\tau $ increases, the erasure region 
covers more area.

\section{Non-degenerate energy levels with a single thermal reservoir}

Now we set $E_1 > 0$ but keep the system connected with a single bath.  The states are non degenerate.  Hence for  any allowed transition,  
heat is exchanged with this bath. For transition between A and B or B and C, $E_1$ energy is exchanged with 
the bath. This changes the internal energy of the demon by the same amount.  For transition
from C0 to A1 (CW rotation), the system absorbs $E$ amount of heat and pull the mass
by doing $w$ amount of work. During this transition the internal energy of the system is increased 
by $2E_1$. Corresponding energy balance equation is 
\begin{equation}
 E=w+2E_1.
\end{equation}

\begin{figure}[!ht]
\vspace{0.5cm}
\begin{center}
\includegraphics[width=15cm]{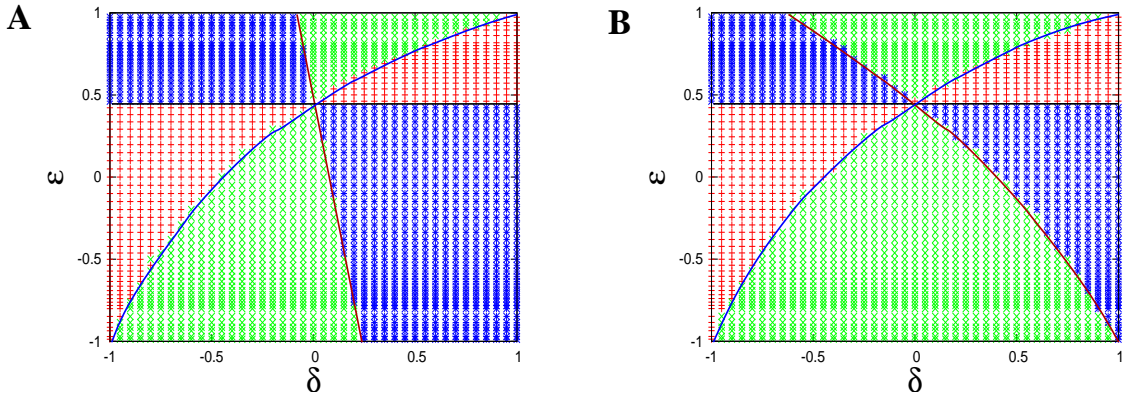}
\caption{ A) Phase diagram when the demon is connected with single bath at temperature T=1.0 for 
$E_1$=0.5 and cycle time $\tau=1.0$. The phase diagram consist of three regions namely
i) Engine (Red plus), ii) Eraser (green cross), iii) Dud (blue star). Here  Line 1, line 2, line 3 
are denoted by black, blue, brown line respectively. B) It represents the phase diagram
when $\tau=10$ and other parameters kept at same value.
}
\label{inf-1}
 
\end{center}
\end{figure}

In Fig.\ref{inf-1}A we have plotted the numerically obtained phase diagram for temperature $T=1.0$ and $E_1=0.5$,
 interaction time is set at $\tau=1.0$. The line 1 now shifted because if we set  $w=0$, it makes $E=2E_1=1.0$
which eventually leads $\varepsilon=0.46$. At line 2 ($\Delta S$=0 and $W=0$) from the earlier condition we get
 $p(0)=\frac{e^{(E-2E_1)/T}}{1+e^{(E-2E_1)/T}}$ and  $p(1)=\frac{1}{1+e^{(E-2E_1)/T}}$. It 
 leads $\delta=tanh((E-2E_1)/2T)<\varepsilon$. As before line 3 ($\Delta S$=0 and $W>0$) 
 meets line 2 at $\delta=0$ which implies $E=2E_1$ and  hence these two lines  meet with line 1 at a single point 
as shown in the Fig.\ref{inf-1}A. As we increase the cycle time ($\tau=10$), line 3 again shifted and the erasure 
region increases (shown in Fig.\ref{inf-1}B).

\section{Non-degenerate energy levels with two thermal reservoirs}

 Let us consider the case when the bath temperatures are different.  For this case we  set $T_c=0.5$ and $T_h=1.0$.
 Note that during each transition the internal energy of the system will change. For 
 each transition couple to the cold bath, internal energy of the demon is changed by
 $E_1$; while internal energy is changed by an amount  $2E_1$ during each transition along  with the hot bath.
 However, in steady state the average internal energy does not change. Hence the demon transfer heat from 
 one bath to the another during its operation. The transfer of energy is only possible if we take
 non-degenerate levels ($E_1 \neq 0$). The main objective of our study is to find the effect of thermal
 bias along with information reservoir and the work reservoir in the performance of the demon. 
 Since $\phi$ represents the average CW rotation, the total work done on the system is  $W=-\phi w$
 while the  heat dissipated to hot  bath is given by 
 \begin{equation}
  Q_h=-\phi E.
 \end{equation}
 As the demon does not accumulate 
 energy on average, using the first law, the heat  dissipated to the cold bath is 
 given by 
 \begin{equation}
  Q_c=\phi 2E_1.
 \end{equation}
 \begin{figure}[!ht]
\vspace{0.5cm}
\begin{center}
 
\includegraphics[width=15cm]{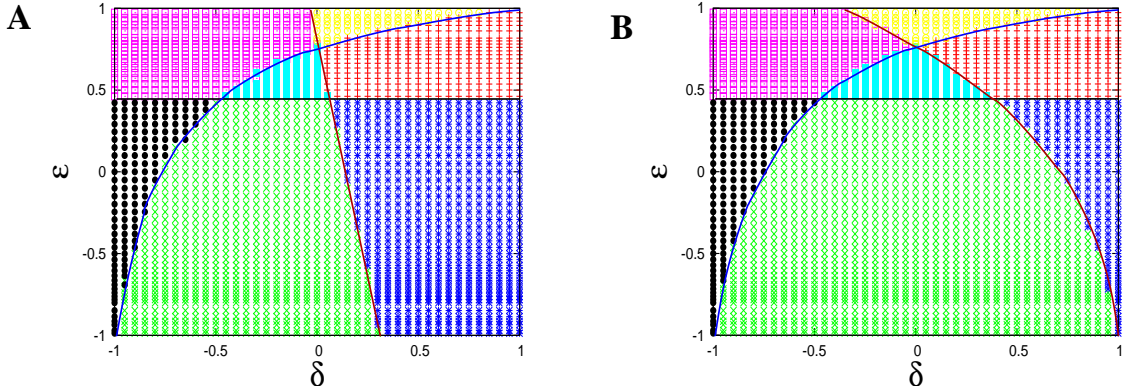}
\caption{ A) Phase diagram when the demon is connected with two baths with  temperature $T_h=1.0$
and $T_c=0.5$. The other parameters set at $E_1$=0.5 and  $\tau=1.0$.
The phase diagram consist of seven regions namely
i)only Engine (Red plus), ii) only Eraser (green cross), iii) Dud (blue star), iv) only refrigerator
(pink open box), v) Eraser and Refrigerator( yellow open circle), vi)Erasure and Engine
(cyan full box), vii) Engine and Refrigerator (black full circle). Here  Line 1, line 2, line 3 
are denoted by black, blue, brown line respectively. B) It represents the phase diagram
when $\tau=10$ and other parameters remains same.
}
\label{phase-1-0.5}

\end{center}
\end{figure}

  As before, the line 1 ($W$=0) is shifted up to $\varepsilon=0.46$. But for this case three lines 
 do not meet at a  point (Fig.\ref{phase-1-0.5}A).  It opens up a triangular area (cyan full box region)
 in the middle. Because from the earlier condition line 2 occurs when 
 $p(0)=\frac{e^{E/T_h -2E_1/T_c}}{1+e^{E/T_h -2E_1/T_c}}$ and $p(1)=\frac{1}{1+e^{E/T_h -2E_1/T_c}}$.
 Now analytical treatment ($\delta=0$) reveals that the crossing point of line 2 and line 3 occurs at 
 $\varepsilon=0.76$ which is consistent with our simulation. Line 1 and line 2 confines the 
 engine region ($W<0$ as denoted by red, cyan, black color); while line 2 and line 3 
 confines erasure region ($\Delta S<0$ as denoted by green, cyan, yellow region) as before. 
 Apart from this, we get a new refrigerator region. The system acts as a refrigerator when
 heat is absorbed from the cold bath i.e., $Q_c<0$. At line 2, not only $W=0$,
 $\Delta S=0$ and $\Delta S_{tot}=0$  but also individually $Q_h=0$ and $ Q_c=0$ (since ($\phi=0$)).
 The left side of line 2 (yellow, pink, black region), 
 $Q_c$ turns out negative and system works as a refrigerator.  Hence apart from four separate regions 
 of engine (red), erasure(green), refrigerator(pink) and dud(blue) there appear three more new 
 and interesting regions in phase diagram where the system performs simultaneously as
 (i) erasure and refrigerator ($\Delta S<0 $ and $Q_c<0$ yellow region),
 (ii) refrigerator and engine ($Q_c<0 $ and $W<0$ black region) even 
 (iii) erasure and engine ($\Delta S<0 $ and $W<0$ cyan region). 
 For the same parameters in Fig.\ref{phase-1-0.5}B we have plotted the phase diagram by increasing the 
 interaction time of the demon with each bit to $\tau=10$. As before the erasure region covers more space. The 
 figures are self explanatory.
 
 \subsection{Discussions:}
 
\begin {figure}[!ht]
 \begin{center}
\includegraphics[width=8 cm]{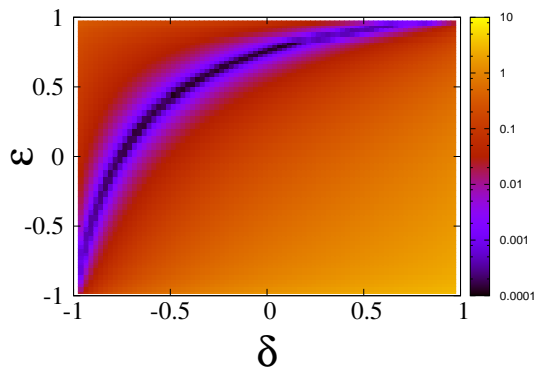}
\caption{ Histogram  of  $\Delta S_{tot}$ for different values of $\delta$ and $\varepsilon$ 
at $T_h$=1.0; $T_c$=0.5,  $E_1$=0.5, $\tau=1.0$}
\label{3d-dist}
\end{center}
\end{figure}

 For a given point in phase diagram  the demon  operates in a steady state.
 The change in entropy of the demon is zero on average. The bath entropy production is given by
 $\Delta S_B=\frac{Q_h}{T_h}+\frac{Q_c}{T_c}$.  The  total entropy production, $\Delta S_{tot}$ will
 be the sum of the entropy change of the memory register $\Delta S$ and the bath entropy production $\Delta S_B$:
 \begin{equation}
 \Delta S_{tot}=\Delta S +\Delta S_B.
 \end{equation}
  It is important to mention that apart from the earlier  two forces (randomizations of bits 
  and pull of gravity), now the demon experiences  temperature difference  between the baths. 
  The phase diagram is a result of interplay of all these three forces.   It is observed $\Delta S_{tot}$ remains always
 greater than zero except at line 2 where it reaches to zero (Fig.\ref{3d-dist}). On this line the demon
 operates in a thermodynamically reversible mode. Hence appearance of all the phases are consistent 
 with the generalized second law of thermodynamics namely $\Delta S_{tot}\geq 0$. In the reversible mode,
 the distributions of bits in the incoming and the  outgoing bit stream are not altered,
 which results all the quantities  $W$, $Q_h$, $Q_c$, $\Delta S$, $\Delta S_B$ 
 individually being zero at line 2.

 In between the line 2 and line 3, $\Delta S<0$ and system behaves as an erasure. However, from 
 the phase diagram (Fig.\ref{phase-1-0.5}A and Fig.\ref{phase-1-0.5}B) we 
 observe that this region is divided into  three separate  parts.  
 At high $\varepsilon$, in between line 2 and 3 (yellow region), the demon acts as an erasure as well as a
 refrigerator. Here $W>0$ and $\Delta S_B>0$. Therefore work is done  on the demon 
 to absorb heat from the cold bath, besides, it erases some information written on the tape.
  In the triangular area in the phase diagram enclosed by the three lines (cyan region), we find   $W<0$,
  furthermore it acts as an erasure.  In this regime even    $\Delta S<0$, the thermal bias plays 
  a dominant role and   makes $\Delta S_B$ more positive so that $\Delta S_{tot}>0$. 
  Note that  $Q_h<0$ and work is extracted on average. In the remaining part( green region) it
  performs as a conventional erasure where work is done on the system to erase the information.

 \begin{figure}[!ht]
 \begin{center}

\includegraphics[width=15cm]{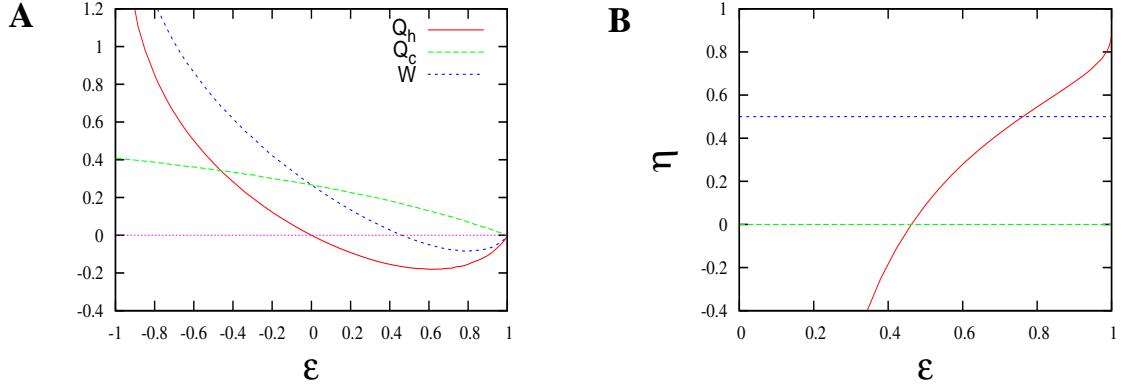}
\caption{A) Variation of $Q_h$, $Q_c$, $W$ with $\varepsilon$ for $\delta$=1.0,
$T_h$=1.0, $T_c$=0.5, $E_1$=0.5 and $\tau=1.0$. 
 B) Variation of efficiency of engine $\eta$ (red line) with $\varepsilon$ while Carnot limit $\eta_c=0.5$ (blue dashed line)}
\label{eta-w}
  
 \end{center}
\end{figure}

 In Fig.\ref{eta-w}A we have plotted $W$, $Q_h$, $Q_c$ as a function of $\varepsilon$ for $\delta=1$. 
 As the every incoming bits are 0, here the demon always write some information into the tape 
 and results $\Delta S  > 0$. We find work can be extracted when $1>\varepsilon>0.46$ and heat is transfered 
 from hot to cold bath. Thus the demon acts as an engine while writing information 
 on the tape.   In the other region ($\varepsilon<0.46 $) work is done on the system as well as
 information is written on the tape and making it dud. Further if we plot the efficiency of the 
 engine defined as $\eta=\frac{W}{Q_h}$, we find $\eta$ can exceed even the Carnot bound 
 $\eta_c=1-\frac{T_c}{T_h}=0.5$ for our case(Fig.\ref{eta-w}B)).  Moreover, in the reversible limit
 at $\varepsilon=1$, $\eta$ tends to a limiting point 1 while both $W$ and $Q_h$ become zero !

  \begin{figure}[!ht]
  \begin{center}
 \includegraphics[width=15cm]{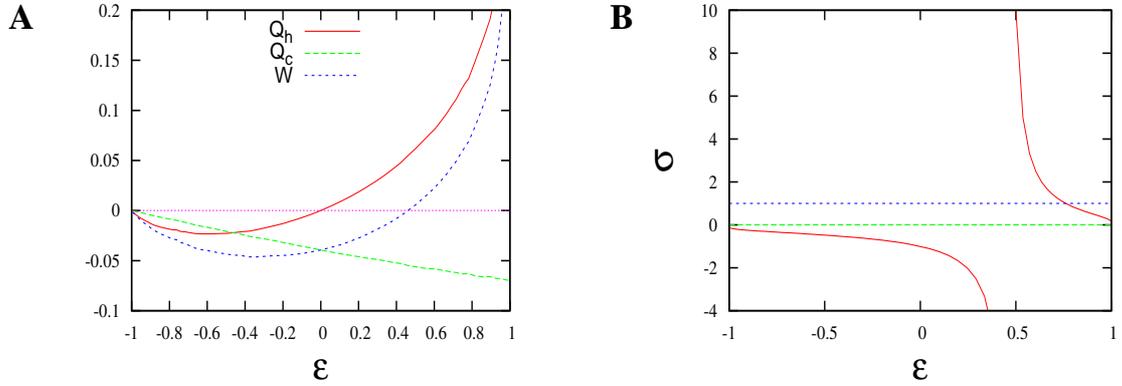}
\caption{ A) Variation of $Q_h$, $Q_c$, $W$ with $\varepsilon$ for 
$\delta$= -1.0, $T_h$=1.0, $T_c$=0.5, $E_1$=0.5 and $\tau=1.0$. 
B) Variation of coefficient of performance for refrigerator $\sigma$ (red line) with $\varepsilon$ 
while Carnot limit $\sigma_c=1.0$ (blue dashed line)
}
\label{cop-w}
  
  \end{center}
\end{figure}

 In Fig.\ref{cop-w}A we have fixed $\delta=-1$ and plotted $W$, $Q_h$ and $Q_c$ vs $\varepsilon$. It
 is observed $Q_c$ always remains negative and act as refrigerator irrespective  of the sign of $W$. For $\varepsilon < 0.46$
 the demon takes heat from the cold bath as well as work is extracted on average. It
  simultaneously acts as a refrigerator and as an engine. This is quite possible because for $\delta=-1$ all 
  the incommoding bits are 1, the outgoing bits will contain a mixture of 0 and 1. The
  system always write information on the tape and $\Delta S  > 0$. Moreover it is observed
  from Fig.\ref{cop-w}B that the coefficient  of performance of the refrigerator   $\sigma=-\frac{Q_c}{W}$,
  is greater than the Carnot limit $\sigma_c =\frac{T_c}{T_h-T_c}=1$ in our case and even it takes 
  negative value  ($\varepsilon<0.46$) as work is extracted while heat is absorbed from cold bath.
  In this respect it can be mentioned that $\phi$ represents the average CW rotation.
 The total work done on the system is  $W=-\phi w$ while the  heat dissipated to hot and cold 
 bath is given by $Q_h=-\phi E$ and $Q_c=\phi 2E_1$ respectively. Then $\eta$, $\sigma$ 
 becomes independent of $\phi$. As we have already fixed $E_1=0.5$ for given $\varepsilon(E)$,
 both $\eta$ and $\sigma$ become constant and they do not depend on $\delta$. Although
 $W$, $Q_h$ and $Q_c$ will depends on $\delta$ because $\phi$ is a function of $\delta$, but,
 their ratio becomes independent of $\delta$ for given $\varepsilon$. 
 
 \section{Conclusion}
 
 We have constructed a simple model of autonomous Maxwell demon which is not 
 manipulated by an external force but it requires a information reservoir (tape) to which it
 can write information. The demon is also connected to two thermal baths. The system 
 reaches unique steady state depending on the model parameters where it 
 exchanges energy from the thermal reservoirs, performs work and write information  on the tape
 at constant rate. The phase diagram shows that the demon can work as an engine by lifting the mass, 
 an eraser by  removing the information written on the tape or a refrigerator by transferring 
 heat from cold bath. Apart from these, we have also found the demon acts simultaneously as
 (i) erasure and refrigerator , (ii) refrigerator and engine, (iii)  erasure and engine.
 We have shown that all this modes of operation are thermodynamically possible
since they do not violate the generalized second law of thermodynamics. 
The efficiency of an engine and coefficient of performance 
of refrigerator can exceed Carnot limit.

\vspace{1cm}
{\large \bf Acknowledgements}
\normalsize

\vspace{0.5cm}
One of us (AMJ) thanks DST, India for financial support. SR thanks  P. S. Pal for useful discussions.



\begin{thebibliography}{10}
\bibitem{max71}  Maxwell J C, Theory of Heat,  1871 Longmans, London.
\bibitem{szi29} Szilard L, On the Decrease of Entropy in a Thermodynamic
System by the Intervention of Intelligent Beings, 1929 Z. Phys. {\bf 53}, 840.
\bibitem{lan61}  Landauer R, Irreversibility and Heat Generation in the Computing Process, 1961
IBM J. Res. Dev. {\bf 5}, 183.
\bibitem{pen70}  Penrose O, Foundations of Statistical Mechanics: A
Deductive Treatment, 1970 Pergamon Press, Oxford).
\bibitem{ben82}  Bennett C H, The thermodynamics of computation - a review, 1982 Int. J. Theor. Phys. {\bf 21}, 905.
\bibitem{ben85}  Bennett C H and  Landauer R, The fundamental physical limits of computation, 1985 Sci. Am. {\bf 253}, 48.

\bibitem{zur89}  Zurek W H, Thermodynamic cost of computation, algorithmic complexity and the information metric, 1989  Nature {\bf 341}, 119.
\bibitem{leff03}  Leff H S and Rex A F,  Maxwell’s Demon 2: Entropy, Classical and Quantum
Information, Computing, 2003 Institute of Physics Publishing, Bristol.
\bibitem{ved09}  Maruyama K,  Nori F, and  Vedral V, Colloquium: The physics of Maxwell's demon and information, 2009  Rev. Mod. Phys. {\bf 81}, 1.
\bibitem{man12}  Mandal D and  Jarzynski C, Work and information processing in a solvable model of Maxwell's demon, 2012 Proceedings of the National Academy of Sciences, {\bf 109}, 11641.
\bibitem{man13} Mandal D,  Quan H T and   Jarzynski C , Maxwell's refrigerator: An exactly solvable model, \zprl{111}{030602}{2013}. 
\bibitem{bar13} Barato A C and  Seifert U, An autonomous and reversible Maxwell's demon, \zepl{101}{60001}{2013}.
\bibitem{bar14} Barato A C and  Seifert U, Unifying three perspectives on information processing in stochastic thermodynamics, \zprl{112}{090601}{2014}.
\bibitem{def13} Deffner S, Information driven current in a quantum Maxwell demon,  \zpre{88}{062128}{2013}.
\bibitem{str13}  Strasberg P,  Schaller G,  Brandes T, and  Esposito M, Thermodynamics of quantum-jump-conditioned feedback control, \zprl{110}{040601}{2013}.
\bibitem{hop14} Hoppenau J and  Engel A, On the energetics of information exchange, \zepl{105}{50002}{2014}.
\bibitem{hor13} Horowitz J M,  Sagawa T, and  Parrondo J M R, Imitating chemical motors with optimal information motors, \zprl{111}{010602}{2013}.
\bibitem{esp12}  Esposito M  and  Schaller G, Stochastic thermodynamics for “Maxwell demon” feedbacks, \zepl{99}{30003}{2012}.
\bibitem{peng16} Peng P and  Duan C, A Maxwell demon model connecting information and thermodynamics,  arXiv:1601.01124.
\bibitem{jar13}  Deffner S and  Jarzynski C, Information processing and the second law of thermodynamics: an inclusive, Hamiltonian approach, 2013 Phys Rev X {\bf 3}, 041003.


\bibitem{sag09} Sagawa T, Ueda M, Minimal energy cost for thermodynamic information processing: measurement and information erasure,  \zprl{102}{250602}{2009}.
\bibitem{par15}   Parrondo J M R,  Horowitz J M and  Sagawa T, Thermodynamics of information, 2015 Nature Physics {\bf 11}, 131.

 

\bibitem{toy10} Toyabe S,  Sagawa T,  Ueda M,  Muneyuki E, and  Sano M, Experimental demonstration of information-to-energy conversion and validation of the generalized Jarzynski equality, 2010 Nature Phys. {\bf 6}, 988.
\bibitem{kos14}  Koski J V,  Maisi V F,  Sagawa T, and  Pekola J P, Experimental observation of the role of mutual information in the nonequilibrium dynamics of a Maxwell demon, 2014 Phys. Rev. Lett. {\bf 113}, 030601.
\bibitem{ber11}  Berut A et al., Experimental verification of Landauer's principle linking information and thermodynamics,  2012 Nature {\bf 483}, 187.

 \end{thebibliography}
\end{document}